\documentclass[aps,amsmath,amssymb,reprint]{revtex4-1}
\usepackage{graphicx} 
\usepackage{graphics} 
\usepackage{dcolumn} 
\usepackage[normalem]{ulem}
\usepackage{cancel}
\usepackage{gensymb}
\DeclareMathOperator\erf{erf}
\draft
\begin{document}
\title{A time-delay model for molecular gas flow into vacuum}
\author{Rajiv Goswami}
\email[Corresponding author. Electronic mail: ]{rajiv@ipr.res.in}
\affiliation{Institute for Plasma Research, Bhat, Gandhinagar -- 382428,
India}
\author{K. A. Jadeja}
\affiliation{Institute for Plasma Research, Bhat, Gandhinagar -- 382428,
India}
\date{\today}
\begin{abstract}
Flow of molecular gas into a complex vacuum system is investigated by a lumped 
parameter model to estimate the time evolution of gas pressure $p_g$, which 
for the first time takes into account the realistic effect of time-delay 
arising due to multiple reasons such as valve response, pumping and transport 
of gas, conductance of the pipe network, etc. The net effect of all such 
delays taken together into a single (constant) delay term gives rise to a 
scalar delay differential equation (DDE). Analytical solutions of a linear 
DDE in presence of external forcing due to the (pulsed) injection of gas and
outgassing from a stainless steel (SS) wall are then derived using the Laplace 
transform method, where the transcendental characteristic equation is 
approximated by means of rational transfer functions such as diagonal Pad{\'e} 
approximation. A good agreement is obtained with the numerical results and it 
is shown from these solutions that a reasonably good match with experimental 
results is obtained only in the presence of a nonzero time-delay. An attempt 
is also made to simplify the DDE into an ordinary differential equation (ODE) 
by using a low-order Taylor series expansion of the time-delay term. A singular 
perturbation method is used to solve the resultant initial value problem (IVP) 
type second-order ODE. It is found that unlike the Laplace transform method, 
the ODE closely approximates the DDE only if the delay is small. Emergence of 
stable periodic oscillations in $p_g$, as a generic supercritical Hopf 
bifurcation, if time-delay exceeds a critical value is then established by 
numerical and Poincar{\'e}-Lindstedt method.
\end{abstract}
\maketitle
\section{\label{one}Introduction and motivation}
It is now widely agreed that there exist a number of gas flow and pressure
measurements that are useful for an efficient operation of many complex systems 
such as, airflows in ventilation systems, helium transport in cryogenic 
devices, gas pipelines, fabrication processes in microelectronics, magnetic 
confinement devices such as tokamaks and stellarators, etc. It is instructive 
to note that the parameter of pressure is also used as an indicator of the 
quality of vacuum present in some of these systems. However, the nature of gas 
flow changes with the gas pressure and so its description is decided by the 
value of the Knudsen number ${\rm{Kn}} = \lambda/d$ ($\lambda$ is the mean 
free path and $d$ is characteristic size of the cross section through which 
the gas is flowing), dividing the flow regions into three main categories, 
viscous, intermediate, and molecular. Although a detailed dynamic model based 
on fluid or kinetic modeling, taking into account several (actual) flow 
interconnections in a complex geometry would be ideally desirable, it may not 
turn out to be easy and fast enough for any real-time objectives. For example,
eventhough molecular flow is theoretically the best understood of any flow 
type,\cite{ohanlonbook} an accurate estimation of conductances would require 
to take into account beaming and exit/entrance corrections, preferably by the
Monte Carlo method. Instead, one could utilise space-averaged models that 
substantially reduce the computational load needed to describe these 
extensively interconnected gas flow networks. This paper presents one such 
model for transient pressure behaviour, and since the experimental results 
needed for its justification have been obtained on a device with a largely 
molecular flow (${\rm{Kn}} > 1$), our theoretical model likewise assumes that 
the flow is molecular throughout the network. This implies that the flow is 
predominantly determined by gas-wall collisions, with diffuse reflection at 
most wall surfaces.\cite{ohanlonbook} A model encompassing the entire range 
of flow, from viscous to molecular, is beyond the scope of this paper.

Many analytical and numerical models exist for molecular flow conditions.
\cite{kendall1969,welch1973,kanazawa1988,ohta1983,wilson1987} Most of these
studies utilise the close correspondence between a molecular flow and an 
electrical circuit. Vacuum conductances are constant and not a function of 
pressure in this flow regime, and thus analogous to electric conductances. It 
is further assumed that conductance through each segment of the network is the 
same as a stand-alone component, and hence a lumped-sum value can be used to 
calculate the network conductance.\cite{theil1994} However, in order to make 
these simple lumped parameter models more realistic, we have included the 
important yet neglected role played by time-delay, which appears to be an 
inherent and non-negligible factor in a large complex system. This time lag 
arising mainly out of the conductances of pipes of assorted shapes and sizes 
connected in series and parallel, is analogous to that of having many serial 
resistors and parallel capacitors in an electrical 
circuit.\cite{poschenrieder1982} Thus, these simpler zero-dimensional (0D) DDE 
models could then provide a promising alternative to the sophisticated 
yet cumbersome partial differential equation (PDE) based fluid/kinetic models, 
especially to quickly estimate some global system parameters such as 
pumping speeds, overall conductance of the pipe network, magnitude of 
outgassing/wall pumping, etc.

As opposed to ODEs, DDEs belong to a class of functional differential 
equations (FDEs) which are infinite-dimensional, since the function that 
specifies the initial value requires an infinite number of points to specify 
it. Consequently, DDEs can have oscillatory and chaotic dynamical behaviours.
\cite{beuterbook} In general, these DDEs arise when modelling processes whose 
rate of change of state at any moment of time $t$ is determined not only by 
the present state, but also by past states. \cite{kolmanovskii} The use of 
delay equations to take into account the transport phenomena in a network of 
pipes can be advantageous, since there is no need to analyze the complex 
spatial details of the transmission and transport. However, the results 
obtained on the basis of this model are only qualitative, because of the 
approximate nature of the hypothesis as well as the low reliability of the 
empirical constants used when deriving these equations. A few questions 
concerning the influence of delays can also be raised, such as (i) can they 
destabilize an otherwise stable linear system, (ii) can they stabilize an 
otherwise unstable system, or (iii) is switching from stability to instablility 
and back to stability possible, etc.\cite{cooke1982,gopalsamybook} 
Nevertheless, DDEs have now gained a wide usage in modelling diverse fields 
ranging from population biology, epidemiology, nonlinear optics, economics, 
control of mechanical systems, etc. \cite{erneuxbook}

The paper is organized as follows: A detailed outline of our simple and fast
global 0D model is presented in Sec. \ref{two}. To validate our theoretical
model with an experimental result, we have chosen to apply it to a magnetic
confinement device called tokamak.\cite{wessonbook} The standard method to 
initiate plasma in these machines is to first evacuate the confinement vessel 
to a background (base) pressure of $\lesssim 10^{-7}$ Torr and then open a
gas reservoir valve, usually at a single toroidal location, for a short 
duration to flow in the working gas (hydrogen or deuterium) so that a typical 
(maximum) pre-fill neutral gas pressure $p_g$ of $\sim 10^{-4}$ Torr is 
established in the chamber, which is then ionized by applying a toroidal 
electric field $E$ via transformer action from a central solenoid. Note that 
the vessel remains under continuous pumping throughout. Our theoretical model 
includes for the first time the effect of time-delay to help describe the 
observed time variation of $p_g$. In Section \ref{three} we first present an 
analytical solution of our DDE model using the Laplace transform method 
wherein the time-delay term leads to a transcendental characteristic equation, 
which is solved by using the diagonal Pad\'{e} approximation. We then attempt 
to simplify the infinite-dimensional DDE into a finite-dimensional equation 
by using a low-order Taylor series expansion to approximate the delay term. A 
singular perturbation method\cite{kokotovic1976,kokotovic1984} is to used 
solve the resultant IVP type second-order ODE. Next, we examine the possibility 
of a Hopf type bifurcation which can occur in phase spaces of any nonlinear 
system with dimension $n \geq 2$.\cite{strogatzbook} Apart from a numerical 
solution, the Poincar{\'e}-Lindstedt method is used to analytically figure 
out whether the limit cycle oscillations are stable (supercritical) or 
unstable (subcritical). Section \ref{four} contains the quantitative validation 
of our model derived pre-fill pressure. To do so we first solve the DDE 
numerically using ADITYA tokamak parameters and then compare these solutions 
with the above mentioned analytical and experimentally obtained results. 
Section \ref{five} is devoted to discussion, and conclusions are summarized.
\section{\label{two}Model}
Usually a detailed treatment of gas flow in tubes has been carried out either
with Clausing-type integral equations or by statistical computations based
on Monte Carlo methods which are particularly well suited for complex shapes
and systems of baffles, traps, etc. \cite{steckelmacher1986} In the present 
paper we have however analyzed the response of an evacuated vessel, SS type 
$304$ in particular, to sources and sinks by the following 0D equation for 
the flow of hydrogen gas. Assuming that the velocities are randomized and the 
concept of scalar pressure is valid, 
\begin{equation}
\frac{d}{dt}\left(p_gV\right) = Q - Sp_g
\label{eq1}
\end{equation}
where $p_g$ is the neutral gas pressure, $V$ is volume of the vessel being 
pumped out, throughput $Q$ includes all significant sources of gas, and $S$ 
is the effective pumping speed delivered by all significant sinks. Now, it 
is known that there exist several other types of sources such as normal 
outgassing, permeation, leaks, etc, as well as sinks in addition to the 
system pumps, such as wall pumping and gauge pumping, etc. The process of 
co-deposition, hydrogen trapping, as well as standard surface cleaning 
techniques such as glow discharge cleaning (GDC) facilitate hydrogen loading. 
Typically, an outgassing model involves diffusion of dissolved hydrogen atoms 
through the bulk material to be adsorbed at surface sites and recombination 
of two adsorbed particles which then desorb as hydrogen molecule from the 
surface.\cite{redheadbook} However, these models do not consider the influence 
of hydrogen traps which may comprise less than $0.1\%$ of all lattice sites, 
but can hold almost all of the hydrogen.\cite{berg2014} It is thus amply clear 
that a precise quantitative modelling of the wall sources and sinks is 
nontrivial. However, to qualitatively account for the generic outgassing/wall 
pumping of an SS surface in high vacuum, we simply introduce an additional 
feedback term proportional to the instantaneous gas pressure in Eq. 
(\ref{eq1}). In addition we have also considered gas input through a 
piezoelectric valve of a known throughput and whose voltage waveforms are 
feedforward programmed to fill the tokamak to a preset (desired) pressure. 

On the other hand, the effects of extensive pipe network and interconnections 
such as traps, baffles, and ducts which impede the flow of gas and reduce the 
pumping effect of the system have been lumped and approximated through the 
overall value of the conductance parameter $C$. Interestingly, the conductance 
$C$ of a pipe depends on whether the gas flow is molecular, intermediate,
or viscous. Thus the effective pumping speed $S$ obtained in a chamber, 
connected by the effective conductance $C$ to a pump having a pumping speed 
$S_{\rm{p}}$ as,\cite{rothbook}
\begin{equation}
S = \frac{C S_{\rm{p}}}{C + S_{\rm{p}}}
\label{eq2}
\end{equation}
Note that a solution of Eq. (\ref{eq1}) will depend on the location where 
$p_g$ is measured. We have assumed that heat transfer effects are negligible 
and the gas quickly adjusts to the temperature of the vessel or pumping ducts, 
and is therefore taken as constant. Fortuitously, it also helps in direct 
comparision with the experimental values since the Bayard-Alpert type hot 
cathode ionization gauge (BA-IG) typically deployed under such vacuum 
conditions in principle measure densities and not pressure. 

Note that Eq. (\ref{eq1}) is valid if we assume a continuous, instantaneous 
and perfect mixing throughout the vacuum vessel. But, a realistic consideration 
would yield that the pressure of the gas being registered in a gauge at time 
$t$ will equal the average pressure in the vessel at some earlier instant, say 
$t-\tau$. Here for simplicity we have assumed that the time-delay parameter 
$\tau$ represents a sum of all delays present in the system and is therefore 
taken to be a single positive constant. 

Combining all of these with the equation of state $p_gV = NkT$, where $N = nV$,
$n$ and $T$ are gas density and (constant) gas temperature, respectively, we 
can now reformulate Eq. (\ref{eq1}) as the following DDE,
\begin{equation}
\begin{split}
\frac{dN(t)}{dt} = Q_0\exp\left[-\left(\frac{t - t_0}{\delta}\right)^2\right]
+ a N(t) - \alpha N(t-\tau)
\label{eq3}
\end{split}
\end{equation}
where, $Q_0 = Q/kT$ and $\alpha = S/V$. Eq. (\ref{eq1}) now describes the
time variation of the number of molecules $N(t)$ (of the working gas) contained 
in the vacuum vessel at time $t$. Here, for analytical tractability as well as 
closeness to actual physical system, we have used a Gaussian function to 
describe the piezoelectric valve behaviour and $t_0 \gg \delta > 0$ is assumed
here. The parameters $Q_0$, $a$, and $\alpha$ are related to the maximum 
throughput of the valve, outgassing ($a > 0$) or wall pumping ($a < 0$) rate, 
and pumping speed, respectively, and for simplicity are taken to be constants. 
Interestingly, the presence of a delayed negative feedback term in Eq. 
(\ref{eq3}) stabilizes an otherwise asymptotically unstable system provided,
\cite{gopalsamybook}
\begin{equation}
\alpha > \left(a^2 + \alpha^2 \tau + a \alpha \tau\right)
\label{eqex1}
\end{equation}
In order to obtain a unique solution of Eq. (\ref{eq3}), we have to now 
specify an initial condition in the form of a function $\theta$ for a period 
of time equal to the duration of the time-delay and then seek a function $N$ 
such that $N(t) = \theta(t)$ for $-\tau \leq t \leq 0$, and $N$ satisfies Eq.
(\ref{eq3}) for $t \geq 0$. In the present case we shall take $\theta(t) 
\equiv N_0$, a positive constant associated with the base pressure values.

Incidentally, Eq. (\ref{eq3}) can also be derived as the linearized form of a 
generalized delay-logistic equation of the type\cite{gopalsamybook}
\begin{equation}
\frac{dN(t)}{dt} = N(t)\left[a_1 - a_2 N(t-\tau) + a_3 N(t) \right]
\label{eqex2}
\end{equation}
and have the asymptotic behaviour 
\begin{equation*}
\lim_{t\to\infty}N(t) = N_0 , \hspace*{0.1in}
N_0 = \frac{a_1}{a_2 - a_3}
\end{equation*}
Now, letting $N(t) = N_0 + x(t)$ yields the following variational equation
\begin{equation}
\frac{dx(t)}{dt} = a x(t) - \alpha x(t-\tau) + f\left(x(t),x(t-\tau)\right)
\label{eqex4}
\end{equation}
where
\begin{align*} 
a = a_3 N_0, \hspace*{0.15in} \alpha = a_2 N_0 \\
f\left(x(t),x(t-\tau)\right) = \frac{1}{N_0}
\left[a x^2(t) - \alpha x(t) x(t-\tau)\right]
\end{align*}
Linear approximation of Eq. (\ref{eqex4}) thus nicely approximates our DDE 
model given by Eq. (\ref{eq3}) sans the external gas input through a valve. 
Further, it will be shown that these equations admit Hopf-type bifurcation to 
a periodic solution and a constant steady state becomes unstable for a 
critical value of $\tau$.\cite{gopalsamybook} Note that in absence of external 
forcing, for autonomous monotone feedback systems with delay, a 
Poincar{\'e}-Bendixson type theorem implies that bound solutions converge 
either to an equilibrium or to a periodic orbit, and thus, chaotic 
trajectories are not possible.\cite{mallet1996}
\section{\label{three}Analytical Results}
Although Eq. (\ref{eq3}) can be solved analytically by the `method of steps',
\cite{Driverbook} the calculations become unwieldy if the delay $\tau$ 
is small relative to the interval on which it is desired to determine the
solution. \cite{elsgoltsbook} Hence, we present two different analytical 
methods for obtaining the solution of the DDE, which are later compared with 
numerical and experimental results. For studying the possibility of Hopf
bifurcation we have resorted to the standard Poincar{\'e}-Lindstedt 
perturbation method, which is then verified by numerical computation.
\subsection{\label{three_two}Laplace transform}
In this subsection we shall present another analytical method to solve a
nonhomogeneous DDE like Eq. (\ref{eq3}) i.e., by means of the Laplace 
transform. Thus Eq. (\ref{eq1}) takes the form
\begin{equation}
\begin{split}
\left(s - a + \alpha e^{-s \tau}\right)\int_0^{\infty} e^{-s t} N(t) dt = 
\frac{N_0}{s}\left(s-\alpha + \alpha e^{-s \tau}\right) \\
+ Q_0 \int_0^{\infty}
\exp\left[-st - \left(\frac{t-t_0}{\delta}\right)^2\right]dt
\end{split}
\label{eq9}
\end{equation}
Now assuming that an inversion can be performed, the solution of 
Eq. (\ref{eq1}) can be expressed by means of a contour integral taken
along the vertical line joining the points $b-iT$ and $b+iT$ in the complex
plane, \cite{bellmanbook} so
\begin{equation}
\begin{split}
N(t) = \lim_{T\to\infty} \frac{1}{2\pi i} 
\int_{b-iT}^{b+iT} e^{s t} 
\Biggl\{\frac{N_0}{s} \Bigl[1 - h^{-1}(s)\left(\alpha - a\right)\Bigr] \\
+ Q_0 h^{-1}(s) \int_0^{T}
\exp\left[-st_1 - \left(\frac{t_1-t_0}{\delta}\right)^2\right]dt_1\Biggr\}ds
\end{split}
\label{eq10}
\end{equation}
where the (transcendental) function $h(s) = s - a + \alpha e^{-s \tau}$ is 
called the characteristic function, and the roots of the characteristic 
equation $h(s) = 0$ are called the characteristic roots. Now to evaluate the 
integral on right side of Eq. (\ref{eq10}), one needs to know all the zeros 
of this exponential polynomial in order to calculate the residues contribution.
However, although there are, in general, infinitely many characteristic roots 
of a transcendental equation, according to complex variables theory, since 
$h(s)$ is an entire function, the characteristic equation cannot have an 
infinite number of zeros within any compact set $|s| \leq R$, for any finite 
$R > 0$. Therefore, most of the system's poles go to infinity. \cite{gubook}

Nevertheless, to further reduce the complexity of locating the zeros of the 
characteristic functions in the complex plane, model order reduction by 
treating an infinite-dimensional system like a finite-dimensional one have been 
proposed. Here the main motivation is to approximate delays by means of 
rational transfer functions and generally involves the truncation of some 
infinite series, which can be achieved via the following approximation:
\cite{richard2003}
\begin{equation}
e^{-\tau s} \approx \frac{p(-\tau s)}{p(\tau s)} \nonumber
\end{equation}
where $p \in \mathbb{R}[\tau s]$ is an appropriate polynomial without any
zero in the right half-plane Re $s \geq 0$, such as the  diagonal Pad{\'e} 
approximation,\cite{reusch1988} Laguerre-Fourier and Kautz series,
\cite{makila1999} etc. In this paper we shall use a diagonal Pad{\'e} 
approximant of order $n \geq 3$, as they are known to have optimal accuracy 
and the odd-order method also preserves positivity. \cite{reusch1988} Thus we 
replace the exponential in the characteristic equation by
\begin{equation}
e^z = \frac{P_n(z)}{P_n(-z)}, \hspace*{0.13in}
P_n(z) = \sum_{m=0}^n\frac{n!\,(2n-m)!}{(2n)!\,m!\,(n-m)!} z^m
\label{eq11}
\end{equation}
We then use Eq. (\ref{eq11}) to obtain the dominant roots of the characteristic
equation, as those closest to the imaginary axis correspond to slowly decaying
components which generate the long term response, whereas, far away poles 
correspond to components that decay rapidly. To gain qualitative insights
into the response characteristics of the time-delay system, we shall plot the 
location of the poles and zeros of the transfer function 
$H(s) \equiv h^{-1}(s)$ in the complex $s$-plane, since it is well known that 
the transfer function poles are the roots of the characteristic equation.
\subsection{\label{three_one}Singular perturbation}
Since DDEs, as demonstrated above, are particularly difficult to study 
analytically,\cite{bellmanbook} attempts have also been made to approximate 
them by using a low-order Taylor series expansion of the time-delay term and 
ignoring high order derivatives.\cite{mazanov1974} Here, assuming the 
(constant) delay $\tau$ to be small, we replace $N(t-\tau)$ with the first few 
terms of a Taylor series such that Eq. (\ref{eq3}) becomes,
\begin{equation}
\begin{split}
\varepsilon^2 \frac{d^2 N(t)}{d t^2} + 
\frac{\alpha}{2}\left(1 - 2 \varepsilon\right) \frac{dN(t)}{dt} 
+ \frac{\alpha}{2}\left(\alpha - a\right)N(t) = \\
\left(\frac{\alpha}{2}\right)
Q_0\exp\left[-\left(\frac{t - t_0}{\delta}\right)^2\right]
\end{split}
\label{eq4}
\end{equation}
where $\varepsilon = \alpha \tau/2$ and subject to the initial conditions
$N(0) = N_0$ and $dN(0)/dt = 0$. 

Since the highest derivative of our differential equation (\ref{eq4}) is 
multiplied by a small parameter $\varepsilon^2$ which lowers the order of 
the ODE when $\varepsilon = 0$, it represents a singular perturbation problem. 
Such problems are typically characterized by a boundary-layer which is a 
narrow region where the solution of a differential equation changes rapidly. 
Although the method of matched asymptotic expansions has been successfully 
applied to numerous problems which involve singular perturbations, they are
usually of the boundary value type \cite{Benderbook,nayfehbook}, whereas here 
we shall use a slightly different perturbative concept to obtain the solution 
$N(t)$ of the IVP given by Eq. (\ref{eq4}). Such concepts have been very 
helpful in control and systems theory and optimization of dynamic systems. 
\cite{kokotovic1976,kokotovic1984} 

It is known that in response to an external stimuli, singular perturbations 
cause a multi-time-scale behaviour of dynamic systems characterized by the 
presence of both slow and fast transients. Note that this decomposition
coincides with the asymptotic expansions into reduced (`outer') and 
boundary-layer (`inner') regions. Typically, the reduced model represents the
slowest (average) phenomena which are usually dominant, while the 
boundary-layer models evolve in faster time scales and represent deviations 
from the predicted slow behaviour. \cite{kokotovic1984} This separation of 
time scales also eliminates stiffness difficulties. We shall assume that 
the solution of Eq. (\ref{eq4}) is slowly varying except in an isolated 
section closer to $t = t_0$, and away from this only boundary-layer the 
behaviour of $N(t)$ is characterized by the absence of rapid variation. 

As a step towards a `reduced-order model', we rewrite Eq. (\ref{eq4}) as a 
linear system,
\begin{subequations}
\begin{align}
\frac{dN}{dt} = z \label{eq5a} \\
\varepsilon^2 \frac{dz}{dt} = -\frac{\alpha}{2}\left(1 - 2 \varepsilon\right)z
-\frac{\alpha}{2}\left(\alpha - a\right)N \nonumber \\
+\left(\frac{\alpha}{2}\right)
Q_0\exp\left[-\left(\frac{t - t_0}{\delta}\right)^2\right]
\label{eq5b}
\end{align}
\end{subequations}
subject to the initial values $N(0) = N_0$ and $z(0) = 0$.
On setting $\varepsilon = 0$, Eq. (\ref{eq5b}) degenerates into an algebraic
equation whose root $\bar{z}$ (bar is used to indicate the variables for 
whom $\varepsilon = 0$) is then substituted into Eq. (\ref{eq5a}).
Its solution $\bar{N}(t)$ is then evaluated, while constrained to the 
prescribed initial condition $\bar{N}(0) = N_0$, thus yielding the 
`reduced model'. We then input $\bar{N}(t)$ back into the root equation
to get $\bar{z}(t)$. Now, to understand when these reduced 
solution $\bar{N}$, $\bar{z}$ approximate the original solution $N$, $z$, 
respectively, we first estimate the error $z-\bar{z}$ by defining a 
boundary-layer correction \cite{kokotovic1976}
\begin{equation}
\eta = z + \beta_1 N
\label{eq6}
\end{equation}
where $\beta_1 = \left(\alpha - a\right)/\left(1 - 2 \varepsilon\right)+
m_1 \varepsilon^2$ and $m_1$ is chosen such that substitution of Eq. 
(\ref{eq6}) into Eqs. (\ref{eq5a})-(\ref{eq5b}) separates out the 
$\eta$-subsystem as
\begin{subequations}
\begin{align}
\frac{dN}{dt} = -\beta_1 N + \eta \label{eq7a}\\
\frac{d\eta}{d\tau} + \beta_2 \eta = \left(\frac{\alpha}{2}\right)
Q_0\exp\left[-\varepsilon^4\left(\frac{\tau - \tau_0}{\delta}\right)^2
\right] \label{eq7b}
\end{align}
\end{subequations}
where $\beta_2 = \alpha\left(1 - 2\varepsilon\right)/2-\beta_1\varepsilon^2$. 
Eq. (\ref{eq7b}) is the `fast' part of Eqs. (\ref{eq5a})-(\ref{eq5b}), and 
$\tau = t/\varepsilon^2$ is the `stretched time scale' defined for all 
$\varepsilon \geq 0$. The solution of the `fast' subsystem Eq. (\ref{eq7b}) 
with the initial condition $\eta(0)$ obtained from Eq. (\ref{eq6}) is then 
input to the `slow' subsystem given by Eq. (\ref{eq7a}). Integration of the 
resultant ODE then gives,
\begin{align}
N(t) = 
\left[\frac{N_0}{\beta_1 - \beta_2/\varepsilon^2}\right] \times \nonumber \\
\left[\beta_1\exp\left(-\beta_2 t/\varepsilon^2\right)-
\frac{\beta_2}{\varepsilon^2}\exp\left(-\beta_1 t\right)\right] + \nonumber \\
\left(\frac{\alpha Q_0 \sqrt{\pi} \delta}{4 \varepsilon^2}\right)
\left(\frac{1}{\beta_1-\beta_2/\varepsilon^2}\right) \left\{
\exp\left[\frac{\beta_2^2\delta^2}{4\varepsilon^4}+
\frac{\beta_2 \left(t_0-t\right)}{\varepsilon^2}\right] \times
\right. \nonumber \\ \left.
\left[\erf\left(\frac{\beta_2\delta^2/\varepsilon^2+2t_0}{2\delta}\right)
- \erf\left(\frac{\beta_2\delta^2/\varepsilon^2+2t_0-2t}{2\delta}\right)
\right] + \right. \nonumber \\ \left.
\exp\left[\frac{\beta_1^2\delta^2}{4}+\beta_1\left(t_0-t\right)\right]
\times \right. \nonumber \\ \left.
\left[\erf\left(\frac{\beta_1\delta^2 + 2t_0 - 2t}{2\delta}\right) -
\erf\left(\frac{\beta_1\delta^2 + 2t_0}{2\delta}\right) \right] \right\}
\label{eq8}
\end{align}
where the condition $N(0) = N_0$ is used to evaluate the constant of
integration. It is easily verified that for $\tau = 0$ the solution given 
by Eq. (\ref{eq8}) reduces to that obtained from Eq. (\ref{eq3}) in case
of zero time delay.
\subsection{\label{three_three}Hopf bifurcation}
It is well known that if all the poles of a characteristic equation turn out 
to have negative real parts, then from the stability standpoint the system is
asymptotically stable as all solutions decay to zero under an arbitrary
initial condition. In case of ODEs, a stability analysis based on the
characteristic equation is almost trivial due to the availability of the
Routh-Hurwitz criterion.\cite{gopalsamybook} In our case if we let
$z = s\tau$, $p = a\tau$, $q = -\alpha \tau$, then the characteristic
equation $h(s) = 0$ (refer \ref{three_two}) is equivalent to
\begin{equation}
pe^z + q - z e^z = 0
\label{eq12}
\end{equation}
A necessary and sufficient condition in order that all roots of Eq.
(\ref{eq12}) have negative real parts is that \cite{cooke1982}
\begin{enumerate}
\item[(i)] $p < 1$ and
\item[(ii)] $p < -q < \left(\theta^2 + p^2\right)^{1/2}$
\end{enumerate}
where $\theta$ is the unique root of $\theta = p\tan\theta$,
$0 < \theta < \pi$. Now, if $\alpha > 0$, these stability conditions reduce
to \cite{cooke1982}
\begin{equation}
a\tau < 1, \hspace*{0.13in}
a < |\alpha| < |a^2 + \left(\theta/\tau\right)^2|^{1/2}
\label{eq13}
\end{equation}
Thus, if Eq. (\ref{eq3}) is stable for $\tau = 0$ i.e, $\alpha - a > 0$, then
either it is stable for all $\tau \geq 0$, or else there exists a value
$\tau_0$ such that it is stable for $\tau < \tau_0$ and unstable for
$\tau > \tau_0$.\cite{cooke1982}
From Eq. (\ref{eq13}) it is easy to deduce that the value of $\tau_0$ at which
the trivial solution of Eq. (\ref{eq3}) loses its asymptotic stability is
given by
\begin{equation}
\tau_0 = \frac{\cos^{-1}(a/\alpha)}{\omega_0}
\label{eqex3}
\end{equation}
where, $\omega_0 = \sqrt{\alpha^2 - a^2}$.
This means that at $\tau_0$, the stability transition must correspond to one 
pair of conjugate simple roots $\pm i \omega_0$, $\omega_0 > 0$ on the 
imaginary axis, while all the other roots lie in the open left-half complex 
plane. Thus, as the delay parameter $\tau$ crosses the critical value $\tau_0$, 
a Hopf bifurcation emerges which leads to the appearance, from the equilibrium 
state, of small amplitude periodic oscillations. Note that in addition to 
Eq. (\ref{eqex3}), a transversality condition also has to be simultaneously 
satisfied for the existence of Hopf bifurcation. It dictates that as the 
control (delay) parameter varies uniformly, the characteristic roots must 
cross the imaginary axis with nonzero speed i.e.,
$\frac{d}{d\tau}\Bigl[\rm{Re}(s(\tau))\Bigr]_{\tau=\tau_0} \neq 0$. 
Implicit differentiation of the characteristic equation then yields
\begin{equation}
\frac{d}{d\tau}\Bigl[\rm{Re}(s(\tau))\Bigr]_{\tau=\tau_0} =
\frac{\omega_0^2}{\left(1 - a\tau_0\right)^2 + \omega_0^2 \tau_0^2}
\label{eqex5}
\end{equation}
We now aim to study these periodic solutions in the vicinity of a Hopf 
bifurcation of a time-delay system, and ascertain whether the bifurcation is
supercritical or subcritical in nature. Here in the present work, as opposed 
to the quite tedious center manifold reduction technique, we shall use the 
classical Poincar{\'e}-Lindstedt method to deduce the stability of the
limit cycle which is generically born in a Hopf bifurcation.\cite{casal1980}
However, we utilise the center manifold theorem which states that the stability 
of the equilibrium point in the full nonlinear equation is the same as its 
stability when restricted to the flow on the center manifold, and any 
additional equilibrium points or limit cycles which occur in a neighbourhood 
of the given equilibrium point on the center manifold are guaranteed to exist 
in the full nonlinear equations.\cite{carrbook} As this method is applicable
to equations which, upon linearization around an equilibrium point have
eigenvalues with negative or zero real part, we can use the theorem to begin 
with the variational given by Eq. (\ref{eqex4}).

In the neighbourhood of $(x,\tau) = (x_0,\tau_0)$, there exists a one-parameter
family $x(t,\epsilon)$, $\tau(\epsilon)$ of periodic solutions, with 
$x(t,\epsilon) \to x_0$ ; $\tau \to \tau_0$ as $\epsilon \to 0$. Defining 
$y = x/N_0$ and the scaled time variable $T = \omega t$, we now seek a 
periodic solution of Eq. (\ref{eqex4}) of the form
\begin{equation}
\begin{split}
y(T) = 1 + \epsilon y_1(T) + \epsilon^2 y_2(T) + \ldots \\
\tau = \tau_0 + \epsilon^2 \tau_2 + \ldots \\
\omega = \omega_0 + \epsilon^2 \omega_2 + \ldots
\label{eqex6}
\end{split}
\end{equation}
It turns out that $\tau_2 \neq 0$, which is the generic situation, then for
small $\epsilon$, the periodic solution exists either supercritically, when
$\tau_2 {\rm{Re}(s^{\prime}(\tau_0))} > 0$, or else subcritically when
$\tau_2 {\rm{Re}(s^{\prime}(\tau_0))} < 0$. \cite{erneuxbook} Thus, the
determination of $\tau_2$ is of particular interest as the right hand side
of Eq. (\ref{eqex5}) is always $> 0$. Using Eq. (\ref{eqex6}) in 
Eq. (\ref{eqex4}), we get the following first three orders:
\begin{subequations}
\begin{align}
\omega_0 y_1^{\prime} = a y_1(T) - \alpha y_1(T-\omega_0\tau_0) 
\label{eqex7a} \\
\omega_0 y_2^{\prime} = a y_2(T) - \alpha y_2(T-\omega_0\tau_0) 
+ a y_1^2(T) \nonumber \\
- \alpha y_1(T)y_1(T-\omega_0\tau_0) \label{eqex7b} \\
\omega_0 y_3^{\prime} = a y_3(T) - \alpha y_3(T-\omega_0\tau_0) 
+ F(T,\tau_2,\omega_2) \label{eqex7c}
\end{align}
\end{subequations}
where, prime denotes $d/dT$ and
\begin{align*}
F(T,\tau_2,\omega_2) = 2 a y_1(T)y_2(T) -\omega_2 y_1^{\prime}(T) \\
- \alpha\Bigl[y_1(T)y_2(T-\omega_0\tau_0) + y_2(T)y_1(T-\omega_0\tau_0)\Bigr] \\
+ \alpha\left(\omega_2 \tau_0 + \omega_0\tau_2\right)
y_1^{\prime}(T-\omega_0\tau_0)
\end{align*}
It can be easily verified that Eq. (\ref{eqex7a}) admits the periodic solutions
$\cos(T)$ and $\sin(T)$, but for definiteness we choose
\begin{equation}
y_1(T) = \cos(T)
\label{eqex8}
\end{equation}
Next, the solution of Eq. (\ref{eqex7b}) is given by
\begin{equation}
y_2(T) = m_1 \sin(2T) + m_2 \cos(2T)
\label{eqex9}
\end{equation}
where the coefficients $m_1$ and $m_2$ are found using the method of 
undetermined coefficients to be
\begin{subequations}
\begin{align}
m_1 = \frac{\left(\alpha + 2a\right)\omega_0}{2\left(\alpha - a\right)
\left(5 \alpha + 4 a\right)} \label{eqex10a} \\
m_2 = \left(\frac{2 \omega_0}{\alpha + 2 a}\right)m_1 \label{eqex10b} 
\end{align}
\end{subequations}
Using Eqs. (\ref{eqex8})-(\ref{eqex9}), we can now readily find $\tau_2$ and 
$\omega_2$ from the two algebraic equations that arise from employing the 
following solvability conditions,
\begin{subequations}
\begin{align*}
\int_0^{2 \pi} F(T,\tau_2,\omega_2) \cos(T) dT = 0 \\
\int_0^{2 \pi} F(T,\tau_2,\omega_2) \sin(T) dT = 0 
\end{align*}
\end{subequations}
Thus,
\begin{subequations}
\begin{align}
\tau_2 = -\left(\frac{m_1}{2 \omega_0}\right)
\left[\frac{\alpha + 2 a - 3 \alpha^2 \tau_0}{\alpha + 2 a}\right]  \\
\omega_2 = -\left(\frac{3 \alpha^2}{4 \omega_0}\right)m_2
\end{align}
\end{subequations}
\section{\label{four}Results}
The ADITYA tokamak is a medium size air-core device having major radius
$R = 0.75$ m and minor radius $a = 0.25$ m, with a stainless steel wall and 
a fully circular graphite limiter at one toroidal location.
The plasma is generated in a stainless steel vessel, normally evacuated to
a base pressure of about $(0.3 - 1) \times 10^{-7}$ Torr. To achieve better 
wall conditioning, apart from regular hydrogen glow discharge cleaning 
(GDC) the vacuum vessel and limiter are baked to $\gtrsim 100 \degree$C.
Hydrogen is used as the working gas and filled from a large reservoir of $H_2$
at $\sim 1.1 \times 10^3$ Torr through a piezoelectric valve (MaxTek MV-112) 
of (maximum) $500$ sccm (air) throughput to achieve a pre-fill pressure of 
$(0.8 - 1.0) \times 10^{-4}$ Torr. To get a standard plasma discharge, the 
pre-fill gas is typically input as a square pulse through a pulse generator a 
few hundred milliseconds prior to the application of the loop voltage. 

The external gas exhaust system on ADITYA tokamak primarily consists of four 
ultra high vacuum (UHV) pumping lines with two turbomolecular pumps (Pfeiffer 
Hipace 2300) of $\sim 1900$ l/s N$_2$ and two cryopumps (CVI Torr TM 250) of 
$\sim 4000$ l/s N$_2$ pumping capacities. This system is used to pump down the 
torus initially and between shots (as well as to remove the gas load during the
shots). Although the influx rate of the fueling gas can also be estimated from 
the pressure drop in a calibrated volume in the supply lines of the injection
valves, in this paper we have used the maximum specified throughput, asssuming
that the piezoelectrically-controlled valve opens fully for a given pulse.

The neutral gas pressure is measured by a Bayard-Alpert type Ionization 
gauge which is located on one of the pumping lines.
We now calculate the three main system parameters used in solving 
Eq. (\ref{eq3}) i.e., $Q_0$, $\delta$, and $\alpha$. To do so, we first take 
the maximum allowed throughput $Q$ as $\sim 1900$ sccm for H$_2$ to calculate 
the input amount of pre-fill gas particles. We further assume that the gas 
molecules are at room temperature ($T = 300 \degree$K) and hence $Q_0 = Q/kT$ 
corresponds to $\sim 7.73 \times 10^{20}$ particles. The parameter $\delta$ 
relates to the time for which the piezoelectric valve remains open to allow 
input of the pre-fill gas. In this paper we have benchmarked our calculations 
with two different valve opening times $\Delta t_v$ of $16$ ms and $13$ ms. 
In both cases the pulse for opening the valve is applied at same instant of 
time i.e., $256$ ms. Here we have used $\delta = 9.027$ ms and $\delta = 
7.334$ ms so that our Gaussian approximation yields 
$\int_{0}^{\infty}\exp[-((t-t_0)/\delta)^2]dt \approx 16$ ms and $13$ ms, 
respectively. An additional positive offset of few milliseconds (from the 
center of the applied step function like profile) is introduced to estimate 
the value of parameter $t_0$, such that the rise of the Gaussian function 
coincides closely with the sudden switching on of the experimental pulse. 
Hence, $t_0 \simeq 275$ ms is used here. The value of $\alpha$ is estimated 
from Eq. (\ref{eq2}) using the overall conductance of the vacuum vessel and 
pumping lines connected to the respective pumps with above mentioned pumping 
speeds. The effective pumping speed $S$ is thus found to be $\sim 5400$ l/s, 
and so $\alpha$ is $\sim 2.7$ for a total volume $V \sim 2.0$ m$^3$.
Lastly and importantly, to eliminate any undesirable influence of plasma as 
well as the magnetic fields on the IG readings, in this paper we have only 
considered data from experimental shots without the presence of these effects.

\begin{figure*}
\center
\includegraphics[width=\textwidth]{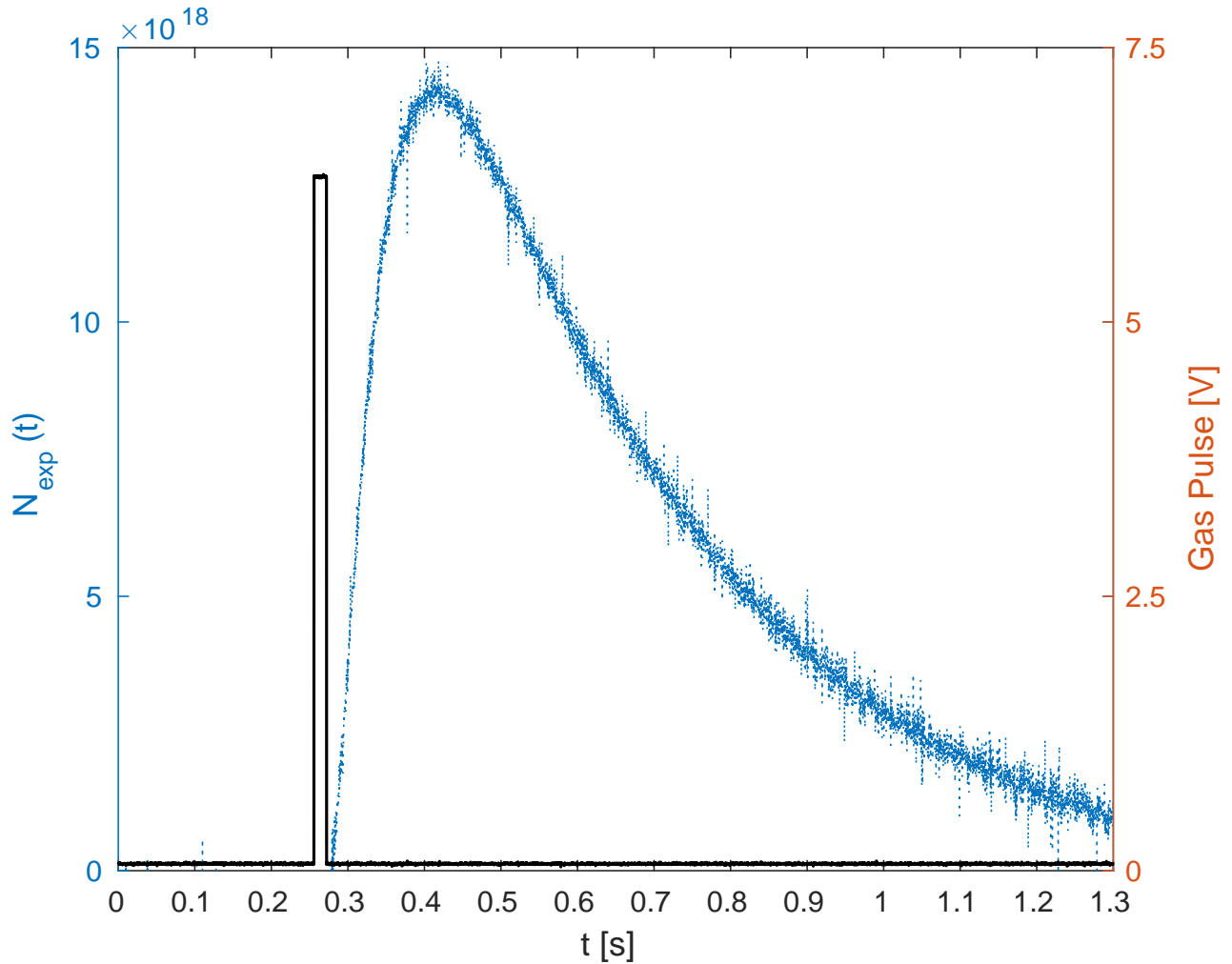}
\caption{Figure showing time variation of the experimentally measured gas 
pressure $p_g$ (converted into number of particles $N$) on y1-axis, with the 
applied gas pulse voltage on y2-axis for shot t1111 on ADITYA tokamak.
Here, the pulse duration is 16 ms and is switched on in a step function
mode at $t \approx 0.256$ s. The BA-IG starts to register the pressure
after a gap of $\sim 23$ ms at $t \approx 0.28$ s.}
\label{fig1}
\end{figure*}

\begin{figure*}
\center
\includegraphics[width=\textwidth]{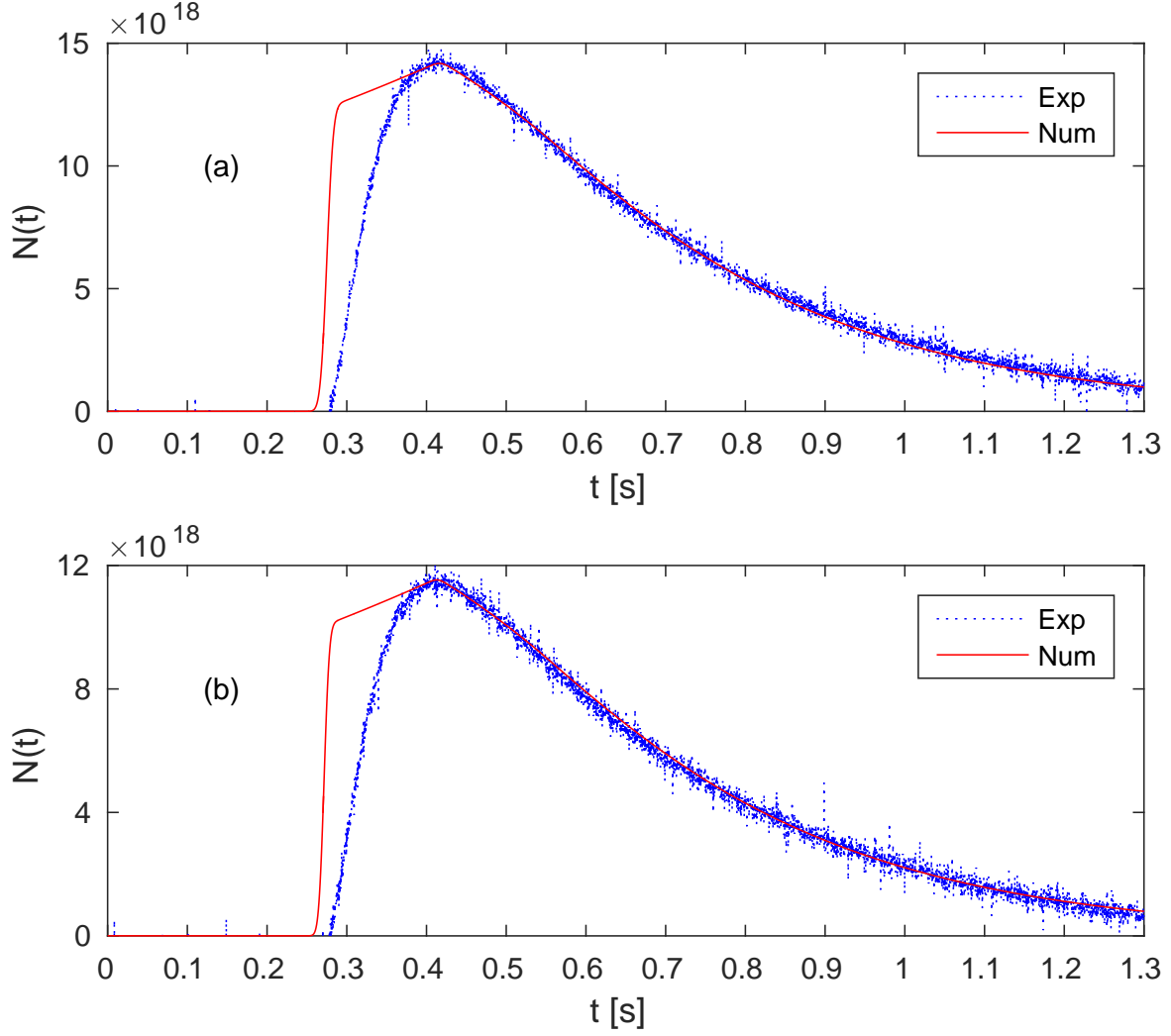}
\caption{Subplots (a) and (b) represent different piezoelectric valve opening 
duration of 16 ms and 13 ms, respectively. The red/solid line represents the 
numerical solution of Eq. (\ref{eq3}) obtained by using the DDE23 solver of 
MATLAB, \cite{matlab} and it is clear that the simple linear DDE model is able 
to reproduce the main features of the respective experimental data shown by 
blue/dotted line. The parameters used for generating the numerical solution
are $\tau = 142.5$ ms, $\alpha = 2.72$, $a = 1.0$, 
$Q_0 = 7.728\times10^{20}$, while $\delta = 9.027 $ ms, $t_0 = 275$ ms for 
subplot (a), and $\delta = 7.334$ ms, $t_0 = 272$ ms for subplot (b), 
respectively.}
\label{fig2}
\end{figure*}

\begin{figure*}
\center
\includegraphics[width=\textwidth]{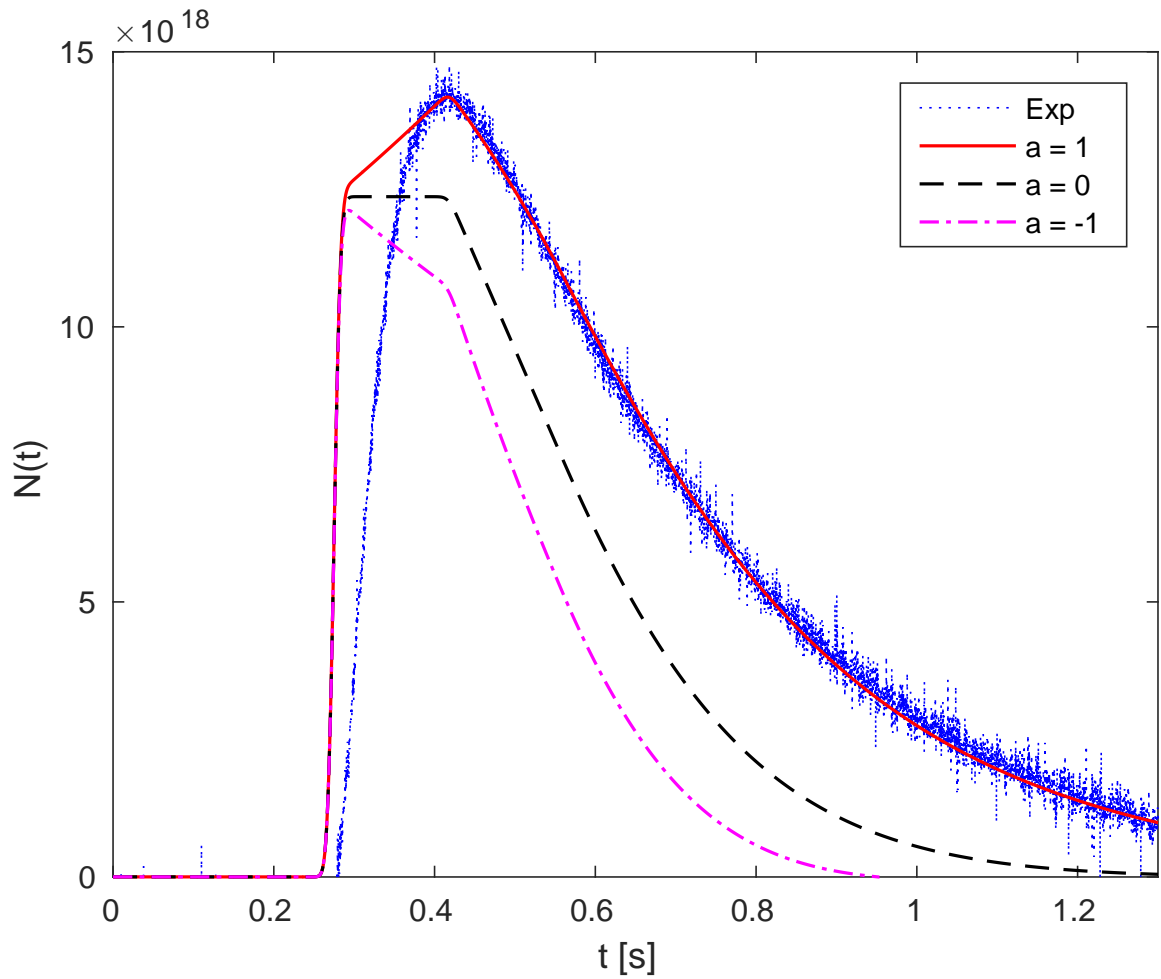}
\caption{To simulate the effect of wall loading/outgassing in Eq. (\ref{eq3}),
change of parameter $a$ while keeping all else as in Fig. \ref{fig2}. 
Here, $a = 1$ (red/solid line) and $a = -1$ (magenta/dash-dot) correspond to 
the wall being a net source and sink of the pre-fill gas, respectively. In 
case of $a = 0$ (black/dash line), the wall plays no role.}
\label{fig3}
\end{figure*}

In Fig. (\ref{fig1}) the y1-axis shows the typical time variation of the total 
number of gas molecules registered by the BA-IG after they had been 
administered few tens of ms before by the piezoelectric valve whose voltage 
pulse is shown on the y2-axis for the same abscissa. For this plot the valve 
remained open for $16$ ms. The two subplots in Fig. (\ref{fig2}) show the 
comparision of the experimental results (blue/dotted line) with the numerical 
solution (red/solid line) of Eq. (\ref{eq3}) obtained by using the DDE23 solver 
of MATLAB.\cite{matlab} Figs. 2(a) and 2(b) correspond to the valve opening 
time of 16 ms and 13 ms, respectively. It is evident that our simple 
first-order linear DDE model with external forcing is able to capture the main 
features of both the experimental profiles. The role of vessel wall acting as
a source or sink of hydrogen gas, decided by the sign of parameter $a$ in 
Eq. (\ref{eq3}), is shown in Fig. (\ref{fig3}). For typical experimental 
parameters, it is deduced from a close match of model solution with $a > 0$ 
(red/solid line) and the experimental data (blue/dotted line), for these 
experiments in particular, that there is a net outgassing.

\begin{figure*}
\center
\includegraphics[width=\textwidth]{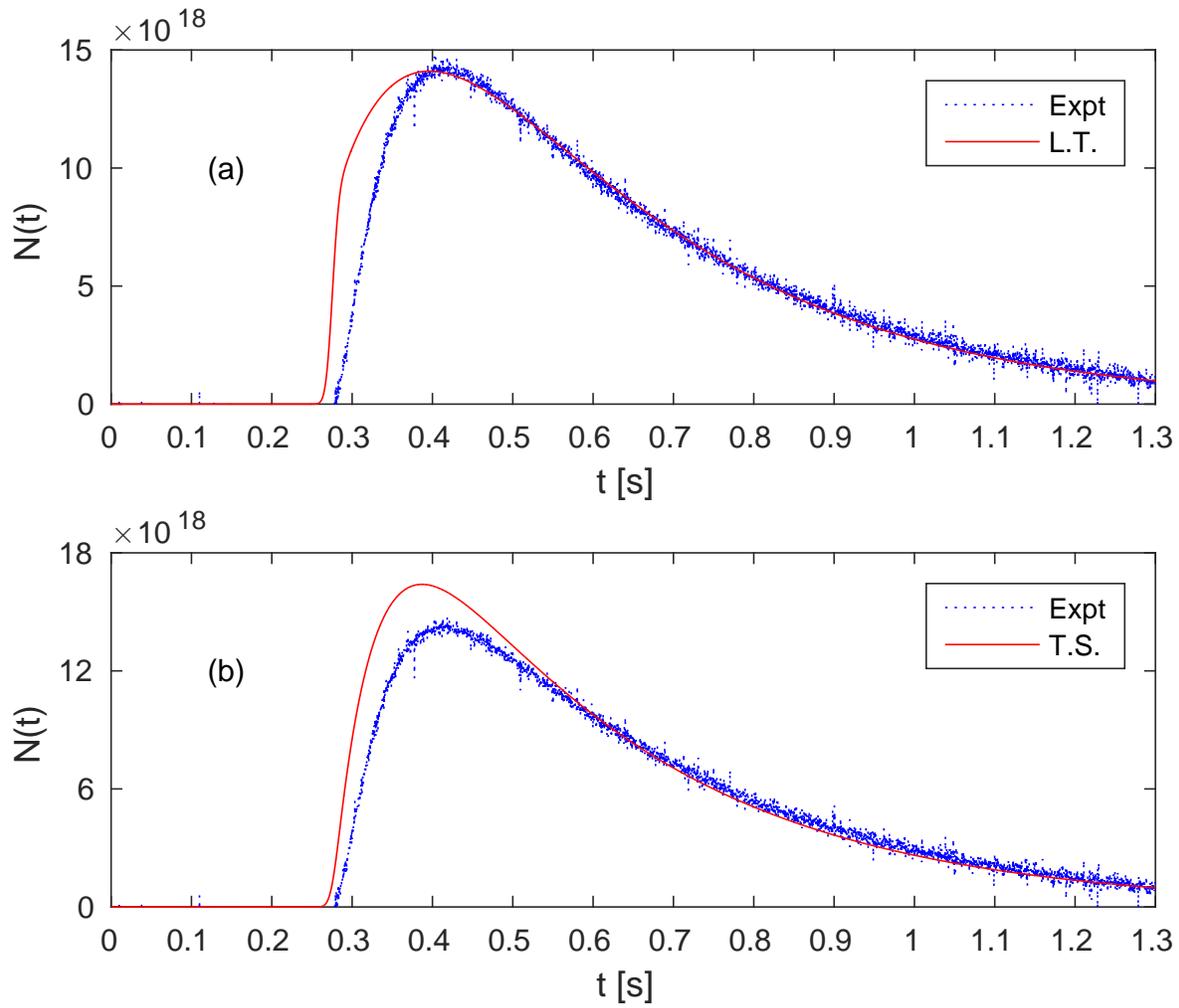}
\caption{Comparision of the analytical solutions (red/solid line) with 
experimental result (blue/dotted line), keeping all the parameters same as in 
Fig. \ref{fig2}(a). Subplot (a) shows the solution obtained  using the Laplace 
transform (L.T.) method (Eq. \ref{eq10}), and subplot (b) shows the solution 
(Eq. \ref{eq8}) obtained on solving the Taylor series (T.S.) expansion by 
singular perturbation technique for an IVP type problem. }
\label{fig4}
\end{figure*}

\begin{figure*}
\center
\includegraphics[width=\textwidth]{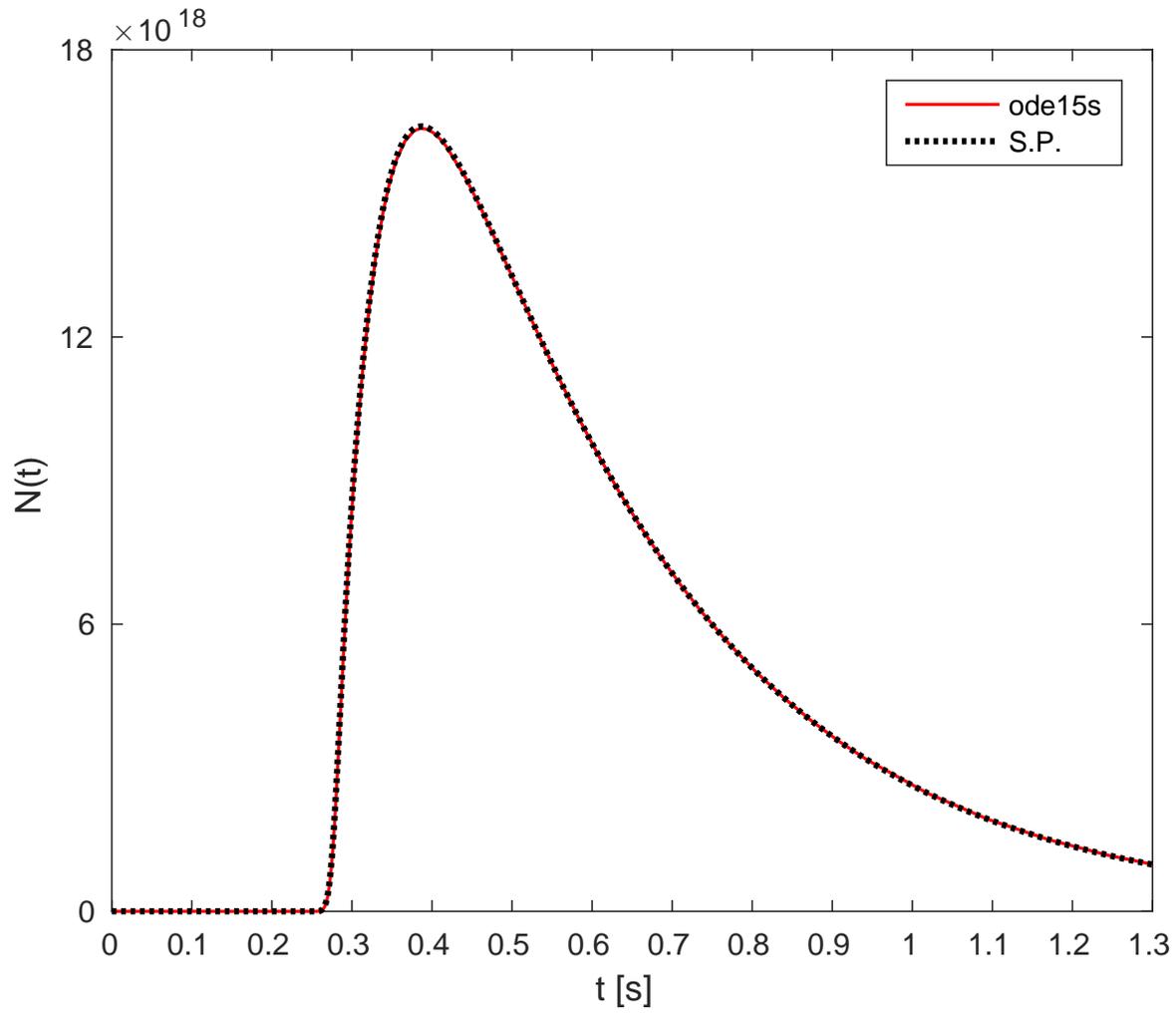}
\caption{Comparision of singular perturbation derived solution given
by Eq. (\ref{eq8}) (dot/black) line with that obtained from Eq. (\ref{eq4}) 
using MATLAB's ode15s solver (solid/red line). An exact match between the two 
solutions establishes the veracity of the analytical technique.}
\label{fig5}
\end{figure*}

In Figs. \ref{fig4}(a) and \ref{fig4}(b), we have plotted the comparisions of
an experimental data ($\Delta t_v = 16$ ms) with the solutions obtained by the 
analytical techniques of Laplace transform and singular perturbation described 
in sections \ref{three_two} and \ref{three_one}, respectively. It is amply 
clear that while the Laplace transform solution much closely approximates the 
experimental data, lending further credence to our model, the singular 
perturbation solution deviates much more. This is because of the conversion of 
a DDE into an ODE by using a low-order Taylor series expansion to approximate 
the time-delay term. It shows that such an approximation is prone to errors.
However, to check the veracity of the singular perturbation solution obtained, 
we have also solved Eq. (\ref{eq4}) using the ode15s solver of MATLAB,
\cite{matlab} and the resultant (complete) match with Eq. (\ref{eq8}) shown in 
Fig. (\ref{fig5}) validates this analytical technique for application to
IVP type singular ODEs.

\begin{figure*}
\center
\includegraphics[width=\textwidth]{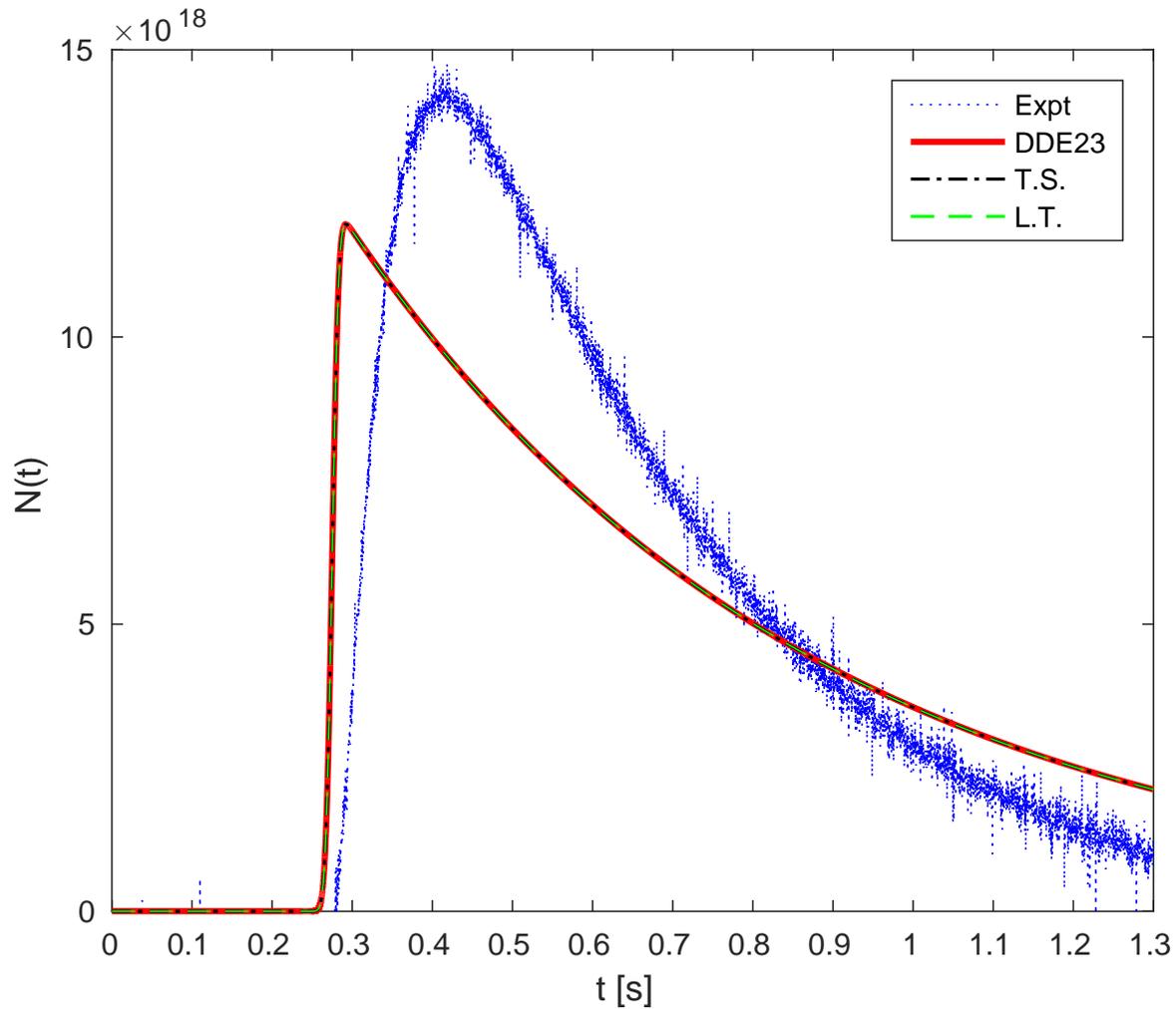}
\caption{Effect of nonexistent time-delay ($\tau = 10^{-9}$), with all 
other parameters being same as in Fig. \ref{fig2}(a). Comparision of solutions
derived using Matlab's DDE23 (red/solid) line, singular perturbation 
(dash-dot/black) line, and Laplace transform with Pad{\'e} approximants 
(green/dash) line is shown. An exact match between the singular perturbation 
and the other two solutions is obtained only in this limit of small $\tau$, 
thus limiting the use of Taylor series expansion for simplifying DDEs. More 
importantly, this plot establishes the crucial role played by a finite 
time-delay.}
\label{fig6}
\end{figure*}

To finally establish the importance of the delayed negative feedback term in
Eq. (\ref{eq3}) and main argument of this paper, we have solved it in the limit 
of a nonexistent delay term ($\tau \approx 0$) with all other parameters being 
the same as in Fig. \ref{fig2}(a). The results (numerical and analytical) are 
shown in Fig. (\ref{fig6}) and it is immediately clear that there is hardly 
any match between the model solutions and the experimental result. This major 
discrepancy shows the necessity of including time-delay in a lumped parameter 
model of the type given by Eq. (\ref{eq1}). Interestingly, it is also seen that 
all the three solutions {\em{viz}}., numerical, Laplace transform, and singular 
perturbation converge exactly with each other. Apart from the robustness of 
the Laplace transform method, this additionally implies that a Taylor series 
expansion of the delay term is suitable to use only if the delay is 
sufficiently small.

\begin{figure*}
\center
\includegraphics[width=\textwidth]{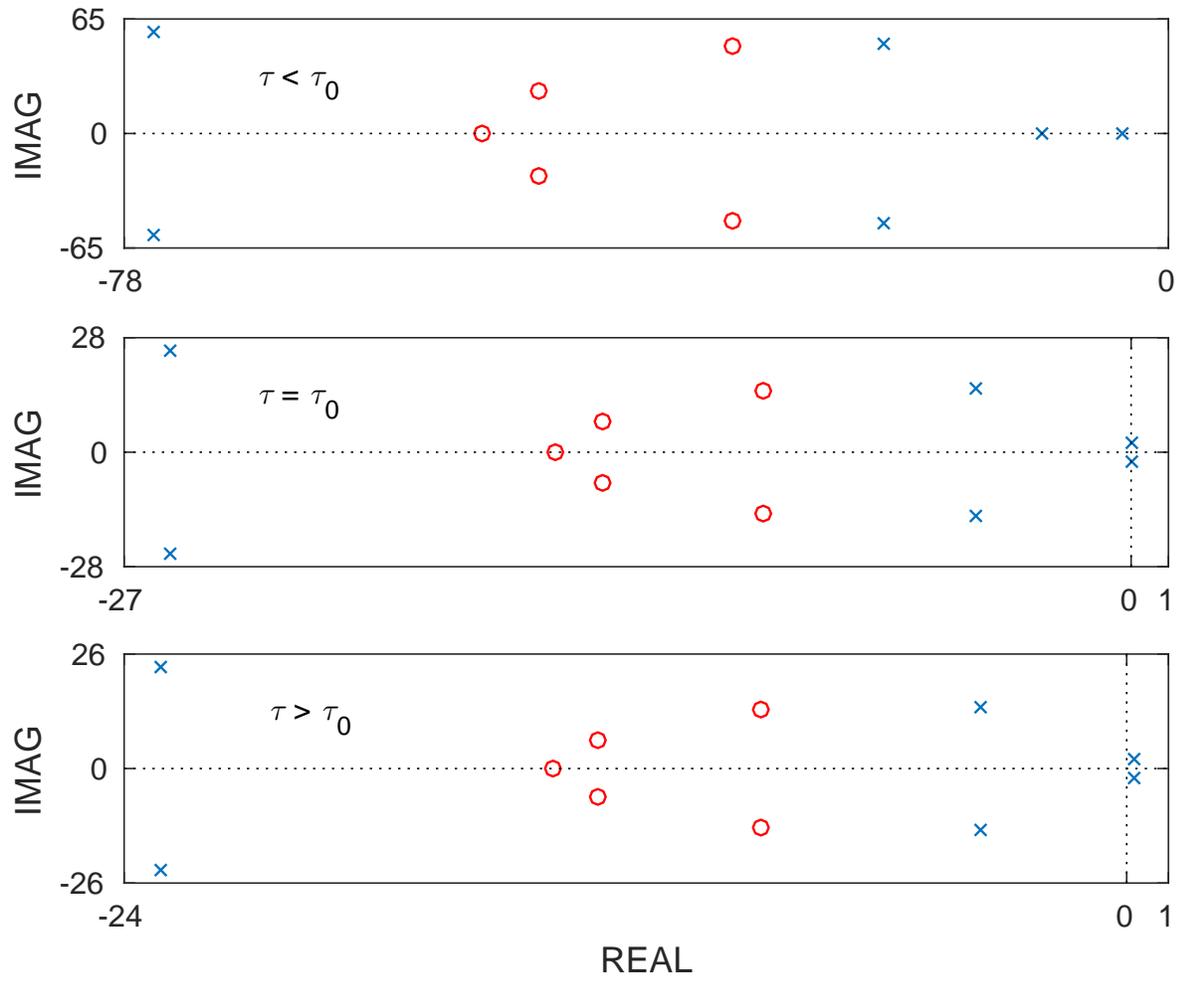}
\caption{Pole-zero plots for three different time-delay values $\tau = 142.5$ 
ms, $\tau = \tau_0 \simeq 472$ ms, and $\tau = 530$ ms, respectively, in the 
complex $s$-plane. The pole locations are marked with a cross ($\times$) and 
the zero locations by a circle ($\circ$). }
\label{fig7}
\end{figure*}

To better understand the Hopf bifurcation described in section 
\ref{three_three}, we first plot the so-called `pole-zero' plots, whose axes 
represent the real and imaginary part of the complex variable $s$. 
It will provide the location of the poles and zeros of the transfer function
$H(s)$, which will then define the system response depending on whether the 
poles are real, imaginary, or complex, occur as single or a conjugate pair, 
and lie in the left-half or right-half of the $s$-plane. For the (fixed)
parameters $a$ and $\alpha$ used in this paper, the critical value $\tau_0$
is $\sim 472$ ms. It can be seen from Fig. \ref{fig7}(a) that for values of
$\tau (< \tau_0)$, all the poles lie on the left side of the imaginary axis
signifying the asymptotic stability of the solutions, while for 
$\tau = \tau_0$ Fig. \ref{fig7}(b) shows, as predicted by the Hopf bifurcation
theorem,\cite{halebook} a pair of purely imaginary roots, whereas Fig. 
\ref{fig7}(c) additionally confirms the presence of Hopf bifurcation, as a 
complex-conjugate pair of eigenvalues cross, at nonzero speed, from left to 
the right half-plane i.e., acquire positive real parts, while the remainder
of the spectrum stays on the left-hand side of the complex $s$-plane.

\begin{figure*}
\center
\includegraphics[width=\textwidth]{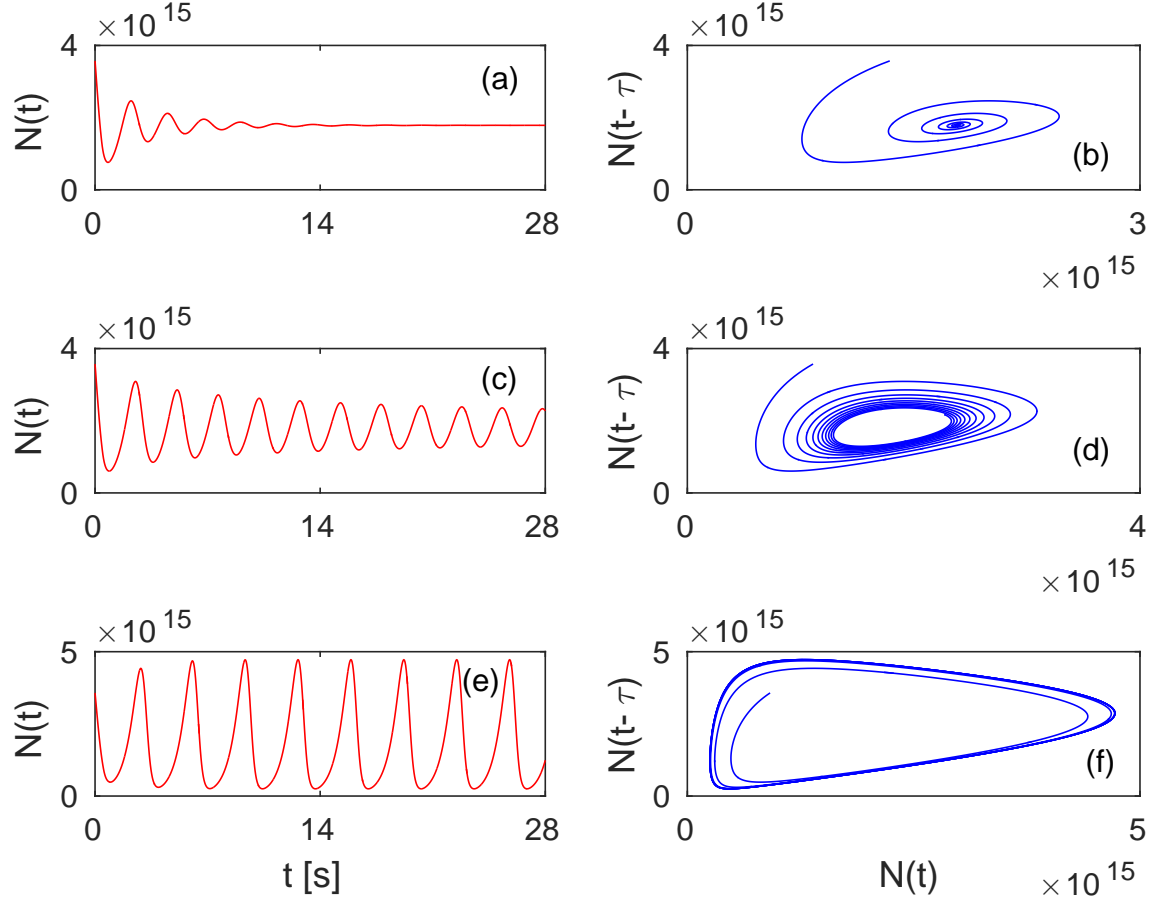}
\caption{Time variation of number density $N(t)$ and phase portraits for three 
different time-delays close to the critical value $\tau_0$; $\tau = 0.41$ for 
subplots (a) and (b), $\tau = \tau_0 \simeq 0.472$ for subplots (c) and (d), 
and $\tau = 0.534$ for subplots (e) and (f), respectively, while keeping all 
other parameters same and as in Figure \ref{fig2}. A supercritical Hopf 
bifurcation with stable periodic orbits is shown to occur for $\tau > \tau_0$.}
\label{fig8}
\end{figure*}

We now numerically examine the stability of solutions in the neighbourhood of 
a Hopf bifurcation characterized by the parameter $\tau_0$ in absence of 
external forcing as given by the nonlinear Eq. (\ref{eqex4}). The presence of
a non T-periodic forcing as represented by the Gaussian will not have an
impact on the solutions as the problem will remain invariant to arbitrary
translations of the origin of time. For $\tau < \tau_0$, Fig. \ref{fig8}(a) 
shows an exponentially damped oscillatory behaviour leading to steady-state 
values, with the orbits spiraling in to a limit point as seen from the phase 
portrait in Fig. \ref{fig8}(b). For $\tau = \tau_0$, the phase portrait in 
Fig. \ref{fig8}(d) shows that the limit point bifurcates into a periodic orbit 
(limit cycle), and Fig. \ref{fig8}(c) displays that the equilibrium state 
loses stability and results in small amplitude oscillations about the steady 
state. Lastly, for $\tau > \tau_0$, Fig. \ref{fig8}(f) shows that the stable 
spiral for $\tau = \tau_0$ changes into an unstable spiral surrounded by a 
small, nearly elliptical limit cycle which becomes distorted as $\tau$ moves 
away from the bifurcation point.\cite{strogatzbook} Fig. \ref{fig8}(e) finally
establishes the supercritical Hopf bifurcation with stable periodic orbits.
In addition, evaluating the analytical condition for our parameters ($a = 1,
\alpha = 2.72$) yields that $\tau_2{\rm{Re}}(s^{\prime}(\tau_0)) \sim 0.18$,  
which is $> 0$. This establishes the bifurcation to be supercritical, thus 
confirming the numerical solution as well.
\section{\label{five}Discussion and Conclusions}
In summary, the new and salient features of our model are that it combines 
the two key mechanisms of delayed negative feedback and external forcing
to yield a lumped parameter model in the form of a linear DDE, which is found 
to reproduce much more accurately the main features of the tokamak pre-fill 
pressure measurements than a reduced model in the form of a delay free system.
This simple, extremely fast, and reasonably accurate model can thus help 
validate not only the experimental findings of the time variation of the 
pre-fill pressure, but also that of the main parameters related to the vacuum 
system such as valve throughput, overall conductance of the pipe network, 
experimental calibration parameters, requisite pumping speeds, outgassing or
pumping by the wall, etc. For example, a detailed calculation of all the 
distributed pumping and conductances of ADITYA tokamak, incorporating 
exit/entrance and beaming corrections for molecular gas flow, will be compared 
with those used in this simple DDE model and published later. Note that in 
tokamak type fusion devices, accurate measurement of pressure has been found 
to be necessary for better estimation of particle balance and ion temperature,
optimization of discharge cleaning, design and testing of vacuum systems, 
accountability of tritium, etc.\cite{dylla1982} 

The model DDE has been solved numerically using a standard solver from MATLAB,
\cite{matlab} and analytically using two different techniques. It is found 
that a singular perturbation technique can be used to solve a second-order 
IVP type ODE that results from approximating the delay term by a low-order 
Taylor series expansion. However, it is also found that such approximations 
work well only if the delay is small. The Laplace transform method on the 
other hand, is highly robust and is able to reproduce the main features of 
the experimental result, but the use of residue theorem during inversion 
requires the solution of an exponential polynomial type characteristic 
equation which can have infinite roots. We have therefore used the diagonal 
Pad{\'e} approximants to solve for the most dominant poles, and the results 
are quite satisfactory when compared with the numerical solution.

Since the DDEs are infinite dimensional, they readily admit the problem of 
bifurcation of steady flow into time-periodic flow, as it is not possible for 
a time-periodic solution to bifurcate from a steady one in one dimension.
\cite{ioossbook} A Hopf bifurcation is shown to occur at a critical value of 
time-delay for which the characteristic equation has exactly one pair of 
conjugate roots on the imaginary axis, and a supercritical bifurcation, where 
the limit cycle is stable above the bifurcation point is found, numerically 
as well as by the Poincar{\'e}-Lindstedt method. Thus the time-delay needs to
be minimized so as to avoid periodic oscillations in $p_g$ which occur if 
$\tau \gtrsim \tau_0$, or else, better wall conditioning can control 
outgassing and lower the value of $a$, which for the same effective pumping 
speed $\alpha$ increases the critical value $\tau_0$. 

Although our simple model is unexpectedly accurate in simulating the pre-fill
pressure profile, further improvement of the model are necessary to help 
understand the mismatch between our model solutions and the experimental data,
especially during the initial rise of pressure. A possible step in that 
direction could be the use of neutral type functional differential equation 
which often arise in the study of two or more simple oscillatory systems with
some interconnections between them,\cite{halebook} instead of the retarded 
type used here.
\nocite{*}
\providecommand{\noopsort}[1]{}\providecommand{\singleletter}[1]{\#1}%
%
\end{document}